\documentstyle{lamuphys}
\makeatletter
\let\chapter\hid@chapter
\makeatother
\begin{document}
\pagenumbering{arabic}
\title{Secular Evolution in Barred Galaxies}

\author{J. A. Sellwood and Victor P. Debattista}

\institute{Rutgers University, Department of Physics \& Astronomy, P O Box
849, Piscataway, NJ 08855, USA}

\maketitle

\bigskip  Rutgers Astrophysics Preprint no 188\footnote{Contribution to {\it
Barred Galaxies and Circumnuclear Activity}, Nobel Symposium {\bf 98}, eds
Sandquist, Aa., Lindblad, P. O. \& J\"ors\"ater, S. (Springer: to appear)}

\begin{abstract}
A strong bar rotating within a massive halo should lose angular momentum to
the halo through dynamical friction, as predicted by Weinberg.  We have
conducted fully self-consistent, numerical simulations of barred galaxy
models with a live halo population and find that bars are indeed braked very
rapidly.  Specifically, we find that the bar slows sufficiently within a few
rotation periods that the distance from the centre to co-rotation is more
than twice the semi-major axis of the bar.  Observational evidence (meagre)
for bar pattern speeds seems to suggest that this ratio typically lies
between 1.2 to 1.5 in real galaxies.  We consider, a number of possible
explanations for this discrepancy between theoretical prediction and
observation, and conclude that no conventional alternative seems able to
account for it.
\end{abstract}

\def\DL{D_{\rm L}}
\def\aB{a_{\rm B}}

\large

\section{Introduction}
Chandrasekhar (1943) showed that a massive object moving though a background
``sea'' of light particles would experience a drag.  The force is the
gravitational attraction by the wake produced by the motion of the massive
object (see e.g., Binney \& Tremaine 1987, \S7.1).  When the mass, $M$, of
the perturber is much larger than that of the individual background
particles, the acceleration takes the form $$
{d v_M \over dt} \propto - \rho \; M \; f\left( {v_M \over \sigma} \right),
\eqno(1)
$$ where $\rho$ is the background density and $f$ is a function of the ratio
of the perturber's velocity, $v_M$, to the (assumed isotropic) velocity
dispersion, $\sigma$, of the background particles.  For a Maxwellian
distribution of velocities, this function is a maximum when
$v_M \simeq 1.37\sigma$.

A similar process must occur as a massive bar rotates inside a halo.  In this
case, the bar creates a wake in the halo which lags the bar and the
gravitational attraction between the bar and the wake produces a torque which
removes angular momentum from the bar and adds it to the halo.  Weinberg
(1985), adapting the perturbation theory approach of Tremaine \& Weinberg
(1984a), estimated the magnitude of the frictional drag force.  Assuming a
massive, rigid bar rotating in an isothermal halo, he concluded that the
spin-down time for the bar would be as little as five bar rotations, for
reasonable parameters!

Early low-quality, but fully self-consistent disc-halo simulations (Sellwood
1980) had previously revealed a rapid loss of angular momentum to the halo
once a bar formed in the disc, at a rate roughly consistent with Weinberg's
prediction.  To our knowledge, no other such simulations have been conducted
in the past 15 years; Combes et al.\ (1990) had only a bulge, not an
extensive halo, of live particles, Raha et al.\ (1991) did not evolve their
models for long enough, the ``halo'' particles of Little \& Carlberg (1991)
were confined to a plane, and the bar used by Hernquist \& Weinberg (1992)
was rigid and their model had no disc.  We here report new fully
self-consistent, simulations superior in many respects to those of Sellwood
(1980), which were designed to reproduce the expected dynamical friction and
to determine the secular changes to the bar, disc and halo in the long term.
Athanassoula (work in progress) is conducting similar experiments using a
direct $N$-body code on a GRAPE device.

\section{New Simulations}

\subsection{Initial Set-up and Numerical Details}
We begin by setting up a disc-halo equilibrium.  The disc particles represent
30\% of the total mass and are laid down with the Kuz'min-Toomre surface
density distribution $$
\Sigma(R) = {Mq \over 2\pi a^2} \left( 1 + {R^2 \over a^2} \right)^{-3/2}.
\eqno(2)
$$ We truncate this profile sharply at the rather small radius of $R=4a$ in
order to limit particle loss from the grid at later times.  We also disperse
the disc particles about the mid-plane in a Gaussian fashion having a uniform
rms thickness of $0.4 a$.

\begin{figure}
\vspace{9.0cm}
\caption{The circular velocity curve at the start of the simulation (solid
curve).  The separate contributions from the disc (dot-dashed curve) and
bulge + halo (dashed curve) are also shown.}
\end{figure}

The remaining 70\% of the mass is represented by the ``halo'' particles, a
dynamically uniform population which could be thought of as comprising both a
luminous bulge and a more extended dark halo.  The halo particles are set in
equilibrium in the manner first used by Raha et al.\ (1991).  They are
selected from an isotropic DF having a (lowered) polytropic form $f =
f[(-E)^m]$, with the limiting energy $E=\Phi_m$, being the potential at some
limiting radius.  The combined disc and halo gravitational potential
distribution to be used in this DF is determined iteratively in the manner
adopted by Prendergast \& Tomer (1970) and Jarvis \& Freeman (1985).  The
polytropic index $m=1.5$ in our case; n.b.\ this corresponds to a standard
$n=3$ polytrope, where $m=n-{3\over2}$ (Binney \& Tremaine 1987).  The
resulting halo mass distribution is not far from spherical and, when combined
with the disc, gives rise to the circular velocity curve in the mid-plane
shown in Figure 1.

Having determined the potential of our initial mass distribution, we set the
disc particles in motion.  Their initial orbits are almost circular, but have
enough random motion to maintain the vertical thickness and to set Toomre's
$Q=0.1$, in the case we focus on here.  We have run other models in which $Q
\geq 1$ at the start and find that the essential results we describe here are
independent of the initial $Q$ value.

Our simulations are performed on a 3-D Cartesian grid having $129^3$ cubic
cells; the code used was described by Sellwood \& Merritt (1994).  We set the
length scale $a=5$ mesh spaces, and chose a time step for the leap-frog
integration of $0.05$ times the dynamical time $\sqrt{a^3/GM}$.  We employ
300K equal mass particles, of which 90K represent the disc.   We adopt units
such that $G=M=a=1$; the rotation period of a particle at the disc half-mass
radius $(R\sim 2.3)$ is about 35 in these units.

\subsection{Evolution}
This model is deliberately designed to be unstable and forms a
strong, rapidly rotating bar within the first 100 dynamical times.  As soon
as the bar forms, a strong torque develops that begins to reduce the total
angular momentum of the disc and to set the halo into rotation, as shown in
Figure 2(a).  Total angular momentum is conserved, of course; almost all that
which is lost from the disc goes into the halo, the tiny remainder being
carried away by escaping particles.

A bi-symmetric distortion is readily detectable in the distribution of halo
particles which initially lags the bar by $\sim 45^\circ$.  As the evolution
proceeds, both the torque and the lag angle gradually decrease, until the
rate of angular momentum transfer ceases almost entirely by $t \simeq 1600$,
at which point the distortion in the halo has become aligned with the bar.

As usual, the bar suffers a bending instability in the early stages (e.g.,
Raha et al.\ 1991).  It is
first detectable at $t \simeq 250$ and is over by $t \simeq 450$.  In this
model, the bar amplitude is not greatly affected by the buckling instability
and the torque on the halo is only slightly reduced by this event.

The pattern speed of the bar also begins to decrease after its formation, as
shown in Figure 2(b).  Figure 2(c) shows that the angular momentum remaining
in the inner part of the disc (where the bar resides) decreases in a similar fashion.  The similar
shapes of these two curves indicates that the bar has a positive moment of
inertia which is approximately, though not exactly, constant, justifying
Weinberg's original assumption.

\begin{figure}
\vspace{17.5cm}
\caption{(a) The time variations of the total angular momenta of the disc and
halo, (b) of the bar pattern speed and (c) of the total angular momentum of
particles in the inner and outer disc.}
\end{figure}

It is interesting that the secular changes seem to end before the halo was
brought to co-rotate with the bar.  Late in the simulation, the mean angular
rotation rate of the halo particles in the very centre is about half that of
the bar, but the rotation of the outer halo is characterised more by a
constant mean orbital speed rather than by uniform rotation.  The alignment
of the halo distortion with the bar appears to indicate that a large fraction
of the halo particles are trapped into resonances with the bar.

As evolution appeared to have almost ceased, we stopped the calculation at
$t=2000$, which corresponds to 40 rotation periods at the initial rotation
rate of the bar.

\begin{figure}
\vspace{9.7cm}
\caption{The time variations of $\DL$ (squares) and of $\aB$ (triangles).}
\end{figure}

Our principal result is displayed in Figure 3, which shows estimates of the
bar length and co-rotation radius at many times during the run.  The
distance, $\DL$, is that from the centre to the Lagrange point on the bar
major axis, and is determined from the potential and pattern speed at each
instant.  The semi-major axis of the bar, $\aB$, is estimated to lie where
the $m=2$ coefficients of a Fourier expansion of the particle distribution
depart from the constant phase and linear fall-off in amplitude
characteristic of the outer bar region.  This definition is consistent
with those adopted by many observers.

The ratio $\DL/\aB$ increases steadily from a value of $\sim 1.3$ at $t=200$,
reaching $\sim 3$ by $t=1200$.  A large value of this ratio is quite unlike
the values believed to pertain in real barred galaxies, as we review next.

\section{Pattern Speeds of Bars in Galaxies}
The only known technique to estimate the pattern speed directly from
observations was proposed by Tremaine \& Weinberg (1984b).  It has been
successfully applied in just one case, the SB0 galaxy NGC~936: Merrifield \&
Kuijken (1995) considerably improved Kent's (1987) original measurement for
this galaxy.  It is to be hoped that this technique will soon be applied to
more galaxies, though it is unlikely to be successful for later Hubble types.

The new spectroscopic and photometric data on NGC~936 yield an estimate of
$\Omega_p \sin i = 3.1 \pm 0.75$ km~s$^{-1}$~arcsec$^{-1}$, which when
combined with Kormendy's (1983) estimated inclination $i=41^\circ$ and
rotation curve, places co-rotation at a distance of $69 \pm 15$ arcsec from
the centre of the galaxy.  This is somewhat outside the visible bar which was
estimated (Kent \& Glaudell 1989) to end at about 50 arcsec from the centre.
For this galaxy, therefore, $\DL/\aB = 1.4 \pm 0.3$.

All other techniques to estimate this ratio in galaxies are indirect.  The
best such evidence comes from the locations and shapes of dust lanes; the
extensive survey of hydrodynamical bar flows by Athanassoula (1992) led her
to conclude that $\DL/\aB = 1.1 \pm 0.1$ would best account for the dust lane
morphology in galaxies.  Her work was, however, restricted to rather
artificial Ferrers bar models, which do not correspond well to the observed
light distributions in bars.  More recently Weiner (1996), P. Lindblad (this
meeting) and others adopt mass models derived from the measured light
distribution; preliminary results suggest that bars are indeed rotating
rapidly, and that $\DL/\aB > 1.7$ seems to be firmly excluded.

There is evidence (Binney et al.\ 1991, Weiner \& Sellwood 1996, Kalnajs this
meeting) that the bar which is believed to reside in the Milky Way also has a
high pattern speed, but the ratio $\DL/\aB$ is not yet firmly established.

Finally, it should be admitted that theoretical prejudice, which was
originally the strongest ``evidence'' for fast bars (see Sellwood \&
Wilkinson 1993), has turned out to be almost worthless, since the bar in our
simulation seems to survive quite happily with a low pattern speed!

While the above evidence could scarcely be described as overwhelming, it
clearly favours values for $\DL/\aB$ that are quite inconsistent with those
we measure from our simulation, at least after the first few bar rotations.

\section{What Could be Wrong?}
One's first reaction to such a puzzling result is to question whether it
needs to be taken seriously.  Of course it is reassuring that the simulation
behaved as theory had already predicted and that similar results are also
being obtained by Athanassoula (1996) using a quite different $N$-body
method.  But perhaps the model differs from real galaxies in respects which
cause it to severely overestimate the importance of dynamical friction, or
that something has been omitted which would counteract the behaviour.

\subsection{Friction Overestimated?}
Chandrasekhar's formula (eq. 1), and Weinberg's analysis, indicates that the
deceleration rate should be proportional to the bar mass and the halo
density.  Could either, or both, of these parameters be too large in our
model?

\subsubsection{Bar Strength}
It is widely believed that bars are massive features, at least in some
galaxies.  They are observed to give rise to strong non-circular motions in
many well studied cases: good examples of strongly non-axisymmetric gas
motions are seen in NGC~5383 (Sancisi, Allen \& Sullivan 1979, Duval \&
Athanassoula 1983), NGC~1365 (J\"ors\"ater \& van Moorsel 1995) and NGC~4123
(Weiner 1996).  A similar streaming pattern in the stellar motions in NGC~936
was observed by Kormendy (1983).  (It should be noted that such streaming
patterns are easily masked in galaxies where the bar lies close to one of the
principal axes of the projected disc.)  Such large non-circular motions seem
to indicate a strongly non-axisymmetric potential, which in turn implies that
the bar has a significant mass compared with the axisymmetric components in
the central parts of these galaxies.

In order to be more quantitative, we have made a comparison between
Kormendy's (1983) slit observations of the barred galaxy in NGC~936 with
similar data taken from our model at $t = 250$ projected and inclined as in
NGC~936.  (This time was chosen because $\DL/\aB$ was close to the the value
of $1.4$ seen in NGC~936.)  We estimate normalized $m=2$ Fourier coefficients
of both the radial and tangential velocities at fixed deprojected radii.
Averaging over a range of radii near the end of the bar, we find we find the
coefficients from our model are 2.8 and 1.7 times larger, for the azimuthal
and radial components respectively, than the same values in NGC~936.  We
conclude that our bar is perhaps twice as strong as that in a typical barred
galaxy, which may therefore have caused us to overestimate the spin down rate
by a factor of two.

\begin{figure}
\vspace{9.1cm}
\caption{The mean specific angular momenta of halo particles at $t=2000$.}
\end{figure}

\subsubsection{Halo Mass}
On the other hand, our halo is nowhere near massive or extensive enough to
give rise to a flat rotation curve beyond the disc edge (Figure 1).  Halo
particles that never come close to it would clearly be unaffected by the bar,
but one would expect the halo out to radii several times the bar semi-major
axis ($\aB \simeq 3$ for our bar) to be torqued up by the bar.  Figure 4
shows the mean specific angular momenta of halo particles at one instant late
in the simulation, plotted as a function of radius, and indicates that
particles far out in the halo have in fact gained disproportionately more
angular momentum than those close in.  Thus if we were to have run a
simulation identical in most respects but having more mass in the outer halo
to give a flat rotation curve, it seems likely that the bar would lose much
more angular momentum.  The under-massive halo of our model therefore gives
too {\it little\/} dynamical friction.

\subsubsection{Halo Core Radius}
Although we do not expect the friction force on a strong bar to behave
precisely as equation (1) would predict, that formula does suggest that the
braking rate should scale with the density and depend on the halo velocity
dispersion.  For an isothermal halo with a core, the velocity dispersion
(assumed isotropic) is largely set by the circular velocity at large radii;
there is therefore little freedom to juggle this parameter for realistic
halos.  On the other hand, a halo having a larger core radius will have a
weaker effect, but only to the extent that friction arises from the inner
halo.

At the cost of eliminating any effective bulge component, we could decrease
the central density of our halo (see Figure 1).  It cannot be decreased
indefinitely, however, since the core radius of a realistic halo cannot be
so large, relative to the disc scale, as to allow the rotation curve to
decline significantly outside the disc.  Thus observed asymptotically flat
rotation curves require a minimum central halo density and a fixed velocity
dispersion at large radii.  Since we have already shown that halo mass at
large radii takes up most of the angular momentum, we do not expect that a
change to the central density, while keeping the halo mass fixed, will affect
friction very much.  Additional experiments to verify this expectation seem
desirable.

\subsubsection{Halo Rotation}
Halos are not expected to have large angular momenta (e.g., Barnes \&
Efstathiou 1987).   We have, nevertheless, tested the possibility that halo
rotation could reduce the bar spin down rate by running two further
simulations, identical in all respects except that in one case some fraction
of the retrograde halo particles had their angular momenta flipped to give a
total halo angular momentum about half the maximum possible.  We found the
bar pattern speed to drop by about the same amount in both and therefore
conclude that dynamical friction is not significantly decreased by giving
the halo even a large positive angular momentum.

\subsection{Effects Omitted?}
\subsubsection{Secondary Bar Growth}
Sellwood (1981) found that when a small bar formed within an extensive disc,
it could grow in length due to trapping of additional stars into the bar as
some angular momentum is removed by spirals in the outer disc.  In his most
extreme case, the bar's half-length approximately doubled from its initial
value.  Since the disc in our simulation was initially truncated at $R=4$ and
the bar which formed had a semi-major axis of fully half the distance to the
initial edge, the scope for significant secondary bar growth is severely
limited in our present model.

Could such substantial bar growth account for the small $\DL/\aB$ ratios of
real galaxies?  We do not think it likely for two main reasons: first, the
bar would have to grow continually which requires incessant spiral activity
in the outer disc.  This is manifestly not happening now in the SB0 galaxy
NGC~936; the bar in this galaxy is likely to have formed some time ago and
with little sign of spiral activity in the outer disc, it cannot have grown
much recently.  Yet a low value of $\DL/\aB$ seems well established in this
particular galaxy.  The other reason is that secondary bar growth makes
the bar longer and stronger, which would increase dynamical friction and
therefore have a less than totally beneficial effect.  Preliminary results
from a further experiment seem to confirm that more extensive discs do not in
fact lead to significantly smaller $\DL/\aB$ ratios.

\subsubsection{Bar Spin-up}
Our simulations are purely stellar and ignore the effects of gas.  It is well
known that the offset shocks on the leading side of the bar cause the gas to
lose angular momentum.  That angular momentum is, of course, given up to the
bar.  However, the amount of angular momentum is quite insignificant, since
the gas mass is already small and the lever arm associated with it is short.

Radial inflows of gas are of slightly greater importance, however.  Increases
in the central mass concentration affect the potential in which the bar
resides, and one consequence is an increase in the bar pattern speed (see also
Kalnajs, this meeting).  In
\S6.1 we give an example in which a substantial mass influx causes the bar
pattern speed to rise by some 25\%.  This is helpful, but on its own, utterly
inadequate to reconcile our simulation with observations.

\section{Assessment}
Thus far we have demonstrated that Weinberg's theoretical prediction of
strong dynamical friction is at least qualitatively confirmed and that the
bar is braked rapidly to an angular rate which is quite inconsistent
with observed $\DL/\aB$ ratios.  We here list the possible solutions to this
discrepancy between theory and observation that have occurred to us or been
suggested by others.

\smallskip
\begin{enumerate}
\item Bars have low pattern speeds

\item Bars are weak

\item Bars grow in length as they slow down

\item Bars are spun up -- e.g., by gas inflow

\item Bars have enormous effective moments of inertia

\item The halo co-rotates with the bar

\item Many halo particles are locked into resonance with the bar

\item Bars do not last long

\item Halos are not very massive
\end{enumerate}
\smallskip

The entire problem hinges on alternative 1 being excluded.  The evidence
for fast bars (\S3) is not as strong as we would wish, and we are
uncomfortable that rather too many of our arguments rest on the assumption
that the early-type SB0 galaxy NGC~936 is typical.  The evidence for massive
halos is strongest for unbarred, late-type spiral galaxies (see \S7).  More
data confirming both a high pattern speed and a massive halo in several
barred galaxies would be most welcome.

We have disposed of possibility 2 and argued that 3 \& 4 are minor
effects that could do little towards removing the discrepancy.  The moment of
inertia of the bar in our simulation, at least, is not large enough to
prevent dynamical friction from slowing it; it seems unlikely that the
structure of real bars is sufficiently different to change this conclusion.
Alternative 6 also does not deserve lengthy consideration -- the angular
momentum of the halo would have to be inconceivably large.

Alternative 7 is somewhat more interesting.  Dynamical friction in our
simulation all but ceases while the bar rotated significantly, which seems to
indicate that many halo particles have become trapped in resonances.  This
phenomenon deserves further investigation, but it is clear that it cannot
provide a solution to our puzzle since friction ceases only after the bar
pattern speed has dropped by a factor of five.

The remaining two alternatives are much more radical, but have to be
contemplated since no other solutions seem tenable.

\section{Transient Bars?}
The possibility that bars could disappear before dynamical friction had
sufficient time to slow them down was first suggested by Hernquist \&
Weinberg (1992).  In order not to violate the bound of $\DL/\aB < 1.7$
suggested by observation, most bars would have to be destroyed quite quickly
-- within 10 rotations, judging from our simulation.  Thus, to maintain the
observed substantial fraction of galaxies containing strong bars (e.g.,
Sellwood \& Wilkinson 1993), this idea requires bars to form and dissolve
more than once over the lifetime of a galaxy.  A second attraction of such a
radical idea is that the fraction of galaxies containing strong bars, for
which there is still no convincing explanation, represents a 30\% duty cycle
in the barred phase.  Regenerating a bar in a disc where one has previously
been destroyed presents a formidable problem, however.

\begin{figure}
\vspace{12.2cm}
\caption{The projected distribution of disc particles at $t=2000$ showing
that a strong, butterfly-shaped bar survives.   We have not included any
bulge/halo particles.}
\end{figure}

\subsection{Bar Destruction}
Many authors have noted that bars are robust, long-lived systems that are not
easily destroyed.  Our simulation provides yet another example; despite
having lost some 2/3 of its angular momentum and having reduced its pattern
speed by a factor of 5, it remains a strong bar, as shown in Figure 5.  Thus
our simulation excludes the possibility, left open by Weinberg's analysis,
that bars simply would not survive such fierce braking.

There are just two known ways to destroy bars: one obvious way is to hit the
bar with a companion, the other is to have a build-up of mass at the bar
centre.  As Athanassoula (this meeting) presents a major study of
bar-satellite interactions, we do not discuss them here.

\begin{figure}
\vspace{8.1cm}
\caption{The time dependence of the bar pattern speed in the 3-D model of
Norman et al.\ (1996).  The non-disc components of this model were rigid but
the bar still slows through interactions with the outer disc until $t=100$.
Mass influx was mimicked by shrinking a rigid Plummer sphere component having
5\% of the disc + bulge mass.  The scale radius of this component decreased
by a factor of 40 over the period $100<t<150$, which caused the pattern speed
to rise before the bar dissolved at $t \simeq 130$.}
\end{figure}

The effects of central mass concentrations have been explored extensively
(Hasan
\& Norman 1990, Hasan, Pfenniger \& Norman 1993, Wada \& Habe 1992, Friedli
\& Benz 1993, 1995, Heller \& Shlosman 1994 and Norman, Sellwood \& Hasan
1996).  The idea here is that the gas driven towards the centre by the bar
itself changes the gravitational potential within the bar to a sufficient
extent that the main orbit family (Contopoulos's family $x_1$) becomes
chaotic, and the regular part of phase space switches to the perpendicular
$x_2$ family. A self-consistent bar can no longer survive once this happens,
and the bar becomes a spheroidal bulge (Norman et al.\ 1996).  The precise
central mass and degree of concentration needed to achieve this has yet to be
firmly tied down, but a few percent of the total mass of the disc seems ample.

Returning for a moment to the point made in \S4.2, we present Figure 6 to
illustrate that the simple process of forming a central mass concentration
increases the bar pattern speed.  This result is taken from the 3-D simulation
by Norman et al.\ in
which the mass build-up was mimicked by simply contracting a rigid spherical
mass component containing 5\% of the disc and bulge mass.  The increase in
pattern speed (by some 25\% in this case) therefore cannot have been caused
by external torques and is simply a result of the changing internal structure
of the bar.

\subsection{Difficulties with Regenerating Bars}
Destruction of a bar, either by the above mechanism or by interaction with a
satellite, would leave the disc in a
dynamically very hot state.  The processes of both its formation and
destruction would disturb the disc stars into quite markedly eccentric
orbits, making the disc quite unresponsive to the kind of large-scale
collective instability needed to reform a bar.  Since only gas can cool, the
galaxy would require a long recuperation period in which a large supply of
fresh gas led to the formation of a substantial fraction of new stars on
nearly circular orbits before the disc could become receptive to a new global
instability.  While the demands here seem excessive, this process could
conceivably occur in the gas-rich late Hubble types; the theory would
therefore appear to predict an increasing bar frequency along the Hubble
sequence that is not observed (Sellwood \& Wilkinson 1993).

Furthermore, if most bars are destroyed by central mass concentrations, the
galaxy will be made more stable.  A high central density is precisely what is
required to stabilize a massive disc (Toomre 1981 and this meeting).  This is
not a watertight argument, since it is not clear that the galaxy would be
absolutely stable no matter how cool the disc, and bars could also be
triggered by interactions (e.g, Noguchi 1987), but it adds considerably to
the difficulties faced by the recurrent bar idea.

Since both methods of bar dissolution make a bulge, bars in bulgeless
galaxies must therefore be their first which, unless they are rotating
slowly, would be required in this picture to be young.  It is perhaps
interesting that most bulgeless barred galaxies are low luminosity galaxies.
Unfortunately, it is unclear what observational data on such galaxies would
be required to test the prediction of dynamical youthfulness.

The idea of transient bars therefore faces extreme challenges.  They would be
reduced somewhat if one could argue that the first bar in a galaxy is braked
by the halo, which is then sufficiently spun-up as to exert much weaker
friction on a second bar.  Some such wildly speculative idea is required if
the regenerated bars alternative is to remain viable.

\section{Low Mass Halos?}
The final possible alternative is that galaxies with fast bars lack massive
halos.  The best evidence for massive halos comes from the extended, flat HI
rotation curves in late-type, unbarred spiral galaxies (e.g., van Albada \&
Sancisi 1986).  Occam's razor, together with current ideas of galaxy
formation, suggest that all galaxies should have flat outer rotation curves,
but the supporting observational data is still sketchy.  Bosma (1992 and this
meeting) concludes that barred galaxies generally do have extensive flat
rotation curves.  An exception for NGC~1365 is claimed by J\"ors\"ater (this
meeting, J\"ors\"ater \& van Moorsel 1995), but deprojection of the complex
kinematic map of this strongly barred, asymmetric, and probably also warped
galaxy is exceedingly difficult.   The evidence for massive halos in
early-type galaxies is also weak because they generally lack the gas disc
which makes such a useful tracer of the potential in late-type systems.  van
Driel and collaborators have attempted to address this issue by mapping the
gas in those rare S0 galaxies that are relatively gas rich, finding some
evidence for flat rotation curve at large radii in the case of NGC~4203 (van
Driel et al.\ 1988).

We therefore think it likely that the circular velocity stays high at large
radii in all galaxies, including those with fast bars.  If this does not
indicate a massive halo, then some alternative explanation for the phenomenon
would need to be invoked (e.g, Milgrom \& Bekenstein 1987).

\section{Conclusions}
The pattern speed problem presented by dynamical friction between a bar and
bulge/halo is
becoming rather insistent.  Most possible solutions seem unattractive, some
are excluded and others need to be stretched excessively.  It is becoming
increasingly difficult to find a tenable conventional explanation.

\bigskip
\noindent {\bf Acknowledgments}  We would like to thank Scott Tremaine for a
critical read of the manuscript.  This work was supported by NSF grant AST
93-18617 and NASA Theory grant NAG 5-2803.

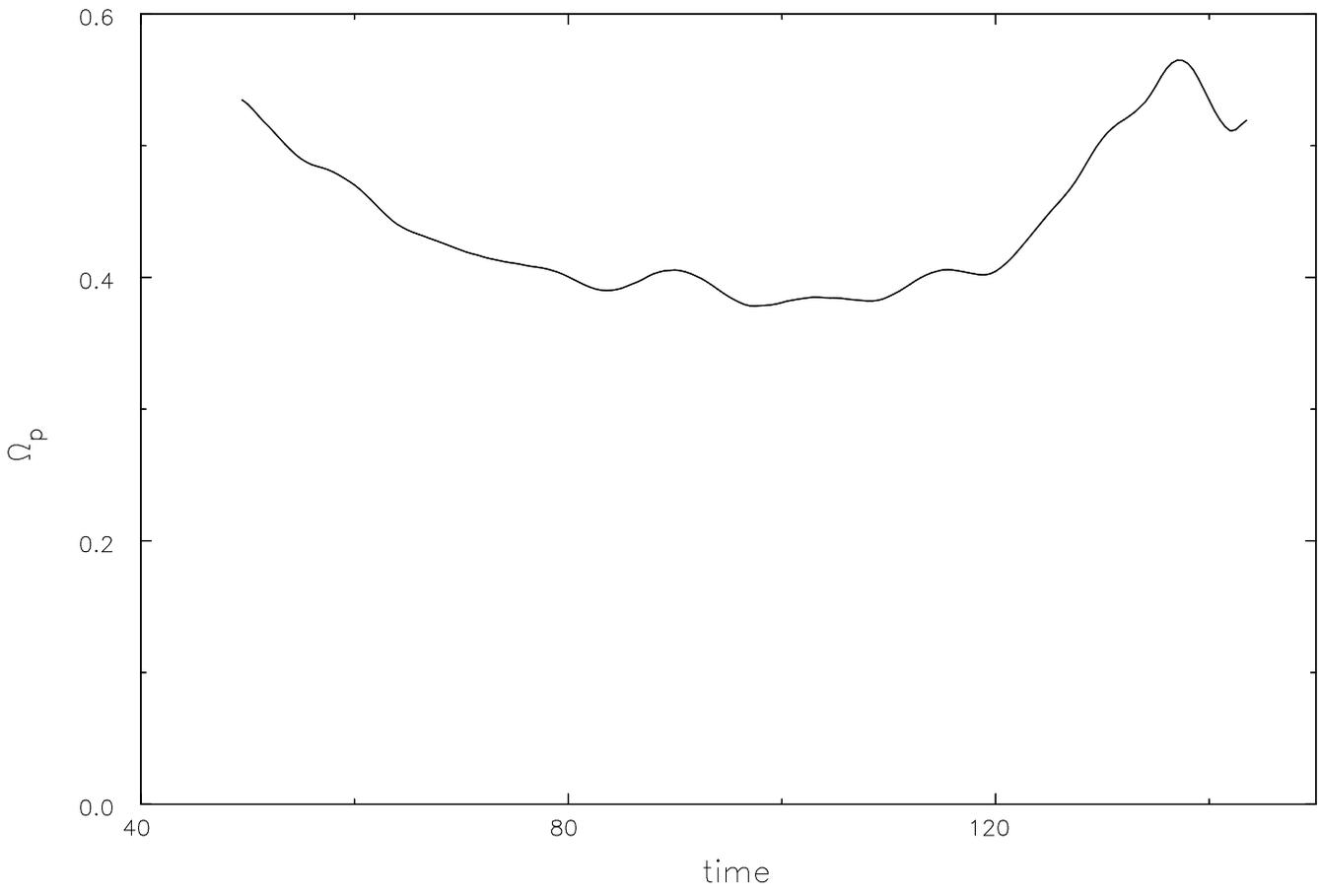
\end{document}